\documentclass[superscriptaddress,twocolumn]{revtex4-1}
\usepackage{graphicx}
\usepackage{amsmath}
\usepackage{amssymb}
\usepackage{braket}
\usepackage{bm}
\usepackage{color}
\usepackage{epstopdf}
\usepackage[stable]{footmisc}
\begin{document}

\title{Quantum steering inequality with tolerance for measurement-setting-errors:
experimentally feasible signature of unbounded violation}

\author{Adam Rutkowski}
\email{fizar@ug.edu.pl}
\affiliation{Institute of Theoretical Physics and Astrophysics, University of Gda\'nsk, ul.~Wita Stwosza 57, 80-952 Gda\'nsk, Poland }
\affiliation{National Quantum Information Center of Gda\'nsk, ul.~W\l{}adysława Andersa 27, 81-824 Sopot, Poland}

\author{Adam Buraczewski}
\email{aburacze@gmail.com}
\affiliation{Institute of Theoretical Physics and Astrophysics, University of Gda\'nsk, ul.~Wita Stwosza 57, 80-952 Gda\'nsk, Poland }

\author{Pawe\l{} Horodecki}
\email{pawel@mif.pg.gda.pl}
\affiliation{Faculty of Applied Physics and Mathematics, Gda\'nsk University of Technology, ul.~Gabriela Narutowicza 11/12, 80-233 Gda\'nsk, Poland}
\affiliation{National Quantum Information Center of Gda\'nsk, ul.~W\l{}adysława Andersa 27, 81-824 Sopot, Poland}

\author{Magdalena Stobi\'nska}
\email{magdalena.stobinska@gmail.com}
\thanks{Corresponding author}
\affiliation{Institute of Theoretical Physics and Astrophysics, University of Gda\'nsk, ul.~Wita Stwosza 57, 80-952 Gda\'nsk, Poland }
\affiliation{Institute of Physics, Polish Academy of Sciences, Al.~Lotnik\'ow 32/46, 02-668 Warsaw, Poland}

\date{\today}

\begin{abstract}
Quantum steering is a relatively simple test for quantumness of correlations, proving that the values of quantum-mechanical measurement outcomes come into being only in the act of measurement. By exploiting quantum correlations Alice can influence -- steer -- Bob's physical system in a way inaccessible in classical world, leading to violation of some inequalities. Demonstrating this and similar quantum effects for systems of increasing size, approaching even the classical limit, is a long-standing challenging problem. Here we provide experimentally feasible signature of unbounded violation of a steering inequality. We derive its universal form where tolerance for measurement-setting-errors is explicitly build-in by means of the Deutsch-Maassen-Uffink uncertainty relation. Then, generalizing the mutual unbiasedness, we apply the inequality to the multi-singlet and multi-particle bipartite Bell-state. However, the method is general and opens the possibility of employing multi-particle bipartite steering for randomness certification and development of quantum technologies, e.g.\ random access codes.
\end{abstract}

\maketitle

In their famous paper, Einstein, Podolsky and Rosen (EPR) highlighted the phenomenon of entanglement~\cite{EPR}: it is possible to see perfect correlations between measurement outcomes obtained by two observers, Alice and Bob, at distant locations, while for each observer his/her outcomes appear to be statistically random. These are the EPR correlations. Validation of entanglement requires designing a specific experimental scenario where measurements on a quantum state give outcomes which violate a classical inequality. The inequality can be constructed, for example, on the basis of probability distribution satisfying the Kolmogorov axioms. Its unbounded violation is equivalent to observation of the EPR correlations which become more and more pronounced when size of a system increases, reaching even classical limit of macroscopic population. This is very challenging to accomplish in the paradigm of Bell-nonlocality testing: if specific observables with $\sqrt{(2^{\log^2d})^d}$ settings and $d$ possible outcomes are used, bipartite quantum states with local Hilbert space dimension $d$ can violate a Bell inequality by a factor of $O\Bigl(\tfrac{\sqrt{d}}{\log^2 d}\Bigr)$~\cite{Junge2010}, later improved to $O\Bigl(\tfrac{\sqrt{d}}{\log d}\Bigr)$~\cite{Junge2011,Buhrman2012,Briet2013}. Thus, an unbounded violation of a Bell inequality requires exponentially many observables (or equivalently, settings). According to the monogamy relation~\cite{Pawlowski2009}, this scaling can be improved only up to the linear one (see the Supplementary Information). However, the present results are still far from this limit and this makes them purely academic as far as the experimental perspective is concerned~\cite{Weiss2016}. 

In case of quantum steering, the task has been found to be less difficult: violation of a steering inequality by a factor of $O\bigl(\sqrt{d}\bigr)$ requires $d+1$ observables in the form of mutually unbiased bases (MUBs)~\cite{Marciniak2015}. However, this scenario necessitates the complementarity relation among the bases to be fulfilled exactly, which is experimentally impossible to attain.

Here we provide a general tight formulation of a quantum steering inequality in which, for the first time, tolerance to the errors of measurement settings is incorporated automatically. This allows for taking a significant step towards feasibility of experimental observation of a signature of unbounded violation of a steering inequality for a multi-particle bipartite Bell-singlet state. In particular, the method enables constructing an inequality which is robust with respect to the degree of unbiasedness of bases and it employs generalized MUBs, where unbiasedness may vary from one observable to another or even form one eigenvector element to another. The inequality is governed by the very parameters involved in the Deutsch-Maassen-Uffink uncertainty relations. Our formula can be adapted to various physical systems but, since mesoscopic quantum effects are very subtle, technological requirements for an experimental apparatus are high. Here we focus on quantum-optical implementation based on polarization entangled squeezed vacuum states generated by parametric-down-conversion sources, polarization rotations and photon counting detection.

Quantum steering was originally introduced by Schr\"odinger in 1935~\cite{Schrodinger1935,Schrodinger1936} and it was elaborated in the course of time~\cite{Gisin1989,Hughston1993}. Recently, it gained a lot of attention because it has been shown to be a weaker test of quantumness of correlations than Bell inequalities thus, less experimentally demanding~\cite{Wiseman2007}. It is different from it in that it quantifies discrepancy between the full quantum-mechanical treatment and the situation where only one of the observers is performing classical measurements (e.g.\ Alice). The latter amounts to the local hidden state model (LHS), where the assumption of objective existence of measurement outcomes without taking any measurements bounds Alice in the possibility of steering Bob's state. However, intriguingly, steering is intimately linked to the Bell nonlocality. In the case of simple CHSH-Bell inequality it is directly limited by the strength of some fine-grained uncertainty relations~\cite{Oppenheim2010}. In a general scenario this relation is more complex~\cite{Ramanathan2015}. In our result we see a connection between multi-particle bipartite steering and the Deutsch--Maassen--Uffink entropic uncertainty relations.

\begin{figure*}\centering
\includegraphics[width=14cm]{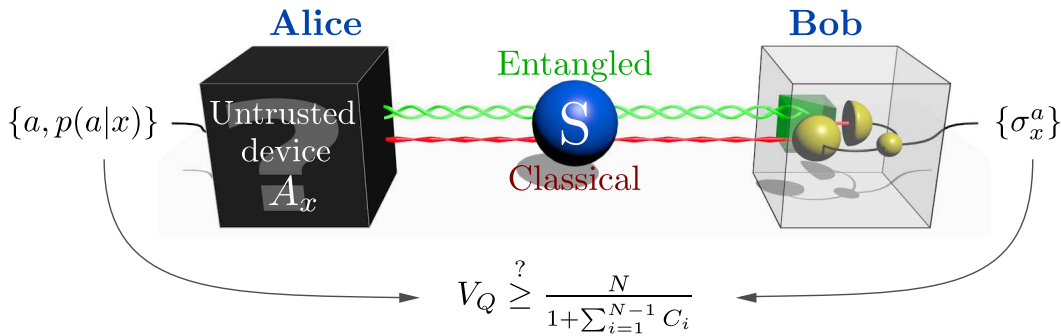}
\caption{Quantum steering scenario: Alice and Bob share a bipartite state $\rho$, either entangled (green) or classical (red), and Bob would like to verify which one this is. Alice performs a measurement using an untrusted device (a black box) of one of her $N$ observables $A_x$, and communicates to Bob the outcome $a$ which she receives with probability $p(a|x)$. Bob applies his measurement $\sigma_x^a$ and checks violation of the steering inequality by computing the lower bound~(\ref{main}). $C_i$ measures maximal complementarity between all Bob's bases, setting at the same time strength of the Deutsch--Maassen--Uffink relations. Quantum-optical implementation assumes a source $S$ of polarized light beams, polarization rotations and photon-number resolved measurements.}
\label{fig:scheme}
\end{figure*}

Violation of steering inequalities has been confirmed in numerous experimental demonstrations involving a single photon~\cite{Wittmann2012,Fuwa2015}, a two-photon singlet~\cite{Smith2012} or a Werner state~\cite{Saunders2010}. Since quantum steering can be formulated as a quantum-information task where the classical measurements simulate an untrusted device, it has been extended to a multipartite scenario useful for semi-device-independent entanglement certification in quantum networks~\cite{Cavalcanti2015}. Our generalization of MUBs will enable developments in the young field of quantum random access codes by allowing for more involved and optimal scenarios than those discussed until now~\cite{Tavakoli2015}.

{\it Main result.} Let us consider a quantum steering scenario shown in Fig.~\ref{fig:scheme}.  Alice and Bob have locally access to subsystems of a bipartite system described by a quantum state $\rho$. Alice chooses one of her settings $x\in\{ 1,\ldots, N\}$, measures a nondegenerate observable $A_x$ with eigenvectors $\{\varphi_{x}^{a}\}$ and receives a result $a \in \{1, \ldots, d\}$ with probability $p(a|x)=\text{Tr}\{(\ket{\varphi_{x}^{a}}\bra{\varphi_{x}^{a}} \otimes I ) \rho\}$. Only after Alice has collected the result $a$, the following conditional state
\begin{equation}
\sigma_{x}^{a}=\text{Tr}_A\{ (\ket{\varphi_{x}^{a}}\bra{\varphi_{x}^{a}} \otimes I) \rho \}
\end{equation}
was ``created at a distance'' at Bob's location. The set $\{\sigma_{x}^{a}\}$ is steerable if for every $a$ Bob can find a basis of $N$ states $\{\phi_{x}^{a}\}$ which is identical to $\{\sigma_{x}^{a}\}$. This is possible only if $\rho$ is maximally entangled. The degree of the match between $\{\phi_{x}^{a}\}$ and $\{\sigma_{x}^{a}\}$ for a given $\rho$ is measured by a steering functional:
\begin{equation}
S_{\text{Q}} = \sup_{\rho \in \mathcal{Q}} \sum_{x=1}^{N} \sum_{a=1}^{d} \text{Tr}\{\ket{\phi_{x}^{a}}\bra{\phi_{x}^{a}} \sigma_{x}^{a}\},
\label{SQ}
\end{equation}
where $\mathcal{Q}$ denotes the set of all bipartite quantum states. The maximal quantum value of (\ref{SQ}) equals the number of settings, $S_{\text{Q}}= N$, and reveals the perfect match.

Within the LHS model Alice's measurements are considered to be untrusted, i.e.\ they are treated as a black box that receives inputs (\hbox{$x$-es}) and reports outcomes (\hbox{$a$-s}). The state $\rho$ is now of a hybrid classical-quantum nature and it determines the form of the Bob's conditional state as an average over ensemble of local hidden states $\sigma_{\lambda}$, which existence is independent of taking any measurements by the observers~\cite{Wiseman2007}:
\begin{equation}
\bar{\sigma}_{x}^{a}=\sum_{\lambda\in\Lambda}q_{\lambda}\,p_{\lambda}(a|x)\,\sigma_{\lambda}.
\label{eq:LHSt}
\end{equation}
The steering functional equals
\begin{equation}
S_{\text{LHS}} = \sup_{\rho \in \mathcal{LHS}} \sum_{x=1}^{N} \sum_{a=1}^{d} \text{Tr}\{\ket{\phi_{x}^{a}}\bra{\phi_{x}^{a}} \bar{\sigma}_{x}^{a} \},
\end{equation}
where $\mathcal{LHS}$ denotes the set of all bipartite states fitting to the LHS model. Its maximal value depends on the choice of Bob's bases $\{\phi_{x}^{a}\}$ and will be estimated below.

Quantum violation of our steering inequality is observed if
\begin{equation}
V_{\text{Q}}=\dfrac{S_{\text{Q}}}{S_{\text{LHS}}} > 1.
\label{VQ}
\end{equation}
We say that it is unbounded if the ratio $V_{\text{Q}}$ between the quantum and classical value of the steering functional is an increasing function of some experimental parameters, for example of the amount of entanglement in $\rho$ or of a number and characteristics of the measured observables. An unbounded violation of~(\ref{VQ}) has been shown for a sequence of maximally entangled states with an increasing local dimension $d$ (e.g.\ maximally entangled states of larger and larger spin)~\cite{Marciniak2015}. However, it necessitates employing very complicated measurements at Bob's side -- the mutually unbiased bases (MUBs) with linearly diverging number of settings $N=d+1$. In general, for a large dimension their existence is known only in same special cases and precise verification of their defining feature that $|\langle \phi_{x}^{a}|\phi_{y}^{b}\rangle|^2=\tfrac{1}{d}$ for $x\neq y$ is experimentally very challenging.

Rather than trying to cure these problems our method shall simply overcome them. To this end, we will relax the MUB condition by assuming near-uniform overlaps and will seek for the maximal overlap between the bases. In this way, our result is naturally linked to the Deutsch--Maassen--Uffink uncertainty relations~\cite{Maassen1988,Deutsch1983} saying that the Shannon entropies of measurement outcomes of two nondegenerate observables obtained in two bases $\{\phi_{x}^{a}\}$ and  $\{\phi_{y}^{b}\}$ for a quantum state $\sigma$ satisfy
\begin{equation}
H \bigl( \{ \phi_{x}^{a} \} \big\vert \sigma \bigr) + H\bigl( \{
\phi_{y}^{b} \}  \big \vert \sigma \bigr) \geq \ -2
\log C_{xy},
\label{M-U}
\end{equation}
where the maximal overlap $C_{xy}=\max_{a,b} \bigl\lvert \braket{\phi_{x}^{a}\vert\phi_{y}^{b}}\bigr\rvert$ is a quantitative measure of complementarity of the bases. The inequality is nontrivial if the two observables do not commute.

Our main result is the following

\begingroup\it
Theorem. Given a quantum steering scenario involving $x\in\{1,\dots,N\}$ settings, $a \in\{1,\dots,d\}$ outcomes, and a set of $N$ orthonormal eingenbases $\left\{ \phi_{x}^{a}\right\}$ defining the receiver's (Bob's) measurements, the LHS steering functional is bounded from above
\begin{equation}
S_{\text{LHS}} \leq 1+\sum_{i=1}^{N-1} C_{i},
\label{SteeringIneq}
\end{equation}
where $C_i=\max_{x }C_{x \,N+x-i}$ and $C_{xy}=\max_{a,b} |\braket{\phi_{x}^{a}\vert\phi_{y}^{b}}|$
for $x,y \in { 1, \ldots, N}$ is defined as in the Deutsch--Maassen--Uffink uncertainty
relations. This implies:
\begin{equation}
V_{\text{Q}} \geq \dfrac{N}{1+\sum_{i=1}^{N-1} C_{i}}.
\label{main}
\end{equation}
In particular, a weaker bound can be derived:
\begin{equation}
V_{\text{Q}} \geq \dfrac{N}{1+ (N-1) C}
\label{quantum-violation}
\end{equation}
with $C=\max_{i} C_{i}=\max_{x\neq y}C_{xy}$.
\endgroup

Let us observe that any state $\rho$ leading to $S_{\text{Q}}(\rho) > 1+ (N-1) C$, where $S_{\text{Q}}(\rho)$ is given by (\ref{SQ}) but without taking supremum over $\mathcal{Q}$, violates the steering inequality~(\ref{SteeringIneq}). The dependence solely on the maximal overlap $C$ frees us from the necessity of a precise control of all MUB conditions separately, as it was done in the approaches taken before ours~\cite{Marciniak2015}. While the physical interpretation of the parameter $C$ is very clear, that of $C_i$ is less comprehensible. $C$ simply quantifies the two most ``compatible'' eigenvectors of the measured observables and equals the maximal element of the $N\times N$ $C_{xy}$ matrix with columns and rows enumerated by the settings indexes $x$ and $y$. Alternatively one may say that $C$ picks the two particular Bob's measurements settings for which the Deutsch--Maassen--Uffink relation exhibits the weakest uncertainty. Interestingly, our mathematical techniques allow us finding the classical bound if we analyze this matrix along its "diagonals" consisting of the elements with indexes $(x, \,y=N+x-i)$, and $C_i$ denotes the maximal element in such "$i$th diagonal".

Now we shall define the relaxed MUB condition by introducing the maximal overlap equal $C \leq \sqrt{d^{\epsilon-1}}$ with $0\le\epsilon<1$, where $\epsilon=0$ corresponds to the original MUB measurements. Any quantum bases satisfying this condition form a generalized MUB. Let us now examine violation of inequality ($\ref{quantum-violation}$) for this $\epsilon$-generalized MUB condition. Unbounded violation is possible whenever the number of settings $N(d)$ is an increasing function of $d$ but, now it can increase arbitrarily slowly. Then the ratio $V_{\text{Q}}\geq C^{-1}$ and it diverges as $\sqrt{d^{1-\epsilon}}$ if $d$ goes to infinity. For $\epsilon=0$ the behavior $V_{\text{Q}}\geq \sqrt{d}$ reproduces the result reported in~\cite{Marciniak2015} but, without the need for the linear scaling $N(d)=d+1$. Equivalently, considering  the dependence on the number of settings by putting $d(N)=N-1$, we obtain at most $V_{\text{Q}}\geq \sqrt{N}$ in the limit of large $N$. However, the best this steering inequality may offer is $V_{Q} \leq N$, which coincides with the maximal possible degree of Bell inequality violation $V_{Q}^{\text{Bell}} \leq N$. Indeed, it follows immediately from the fact that $S_{\text{LHS}} \geq 1$  which is easy to derive by considering Alice's and Bob's measurement statistics on a pure product state (independently on the number of settings and outcomes). Explanation of the discrepancy between the bounds $\sqrt{N}$ and $N$ calls for further research, but it might hold true that the present $\sqrt{N}$ scaling is optimal and addresses the issue of optimality of the MUB settings.

We now would like to turn our abstract mathematical result into a form which could be tested in a laboratory. We first consider a source of independent pairs of photons entangled in their polarizations, i.e. many copies of singlet states $|\Psi \rangle=|\psi_{-} \rangle^{\otimes k}$. We assume a single pair fidelity $F$, efficiency of each detector at Alice side $\eta$, which is required for conclusive version of a steering experiment (see \cite{Smith2012}), and the relaxed MUB condition $C\leq \sqrt{d^{1-\epsilon}}$. In this case the local dimension of the state equals $d=2^{k}$ (allowing existence of $d+1$ original MUBs with $C =\sqrt{d^{-1}}$, see e.g.~\cite{MUb2003}) and we take the number of settings growing slower than the dimension, $N=d^{1-\sigma}$ for $0 \leq \sigma < 1$ ($N$ is assumed to be integer). This leads to the ratio $V_{Q}^{\eta}=\frac{(2^{1-\sigma}\eta F)^{k}}{1+(2^{k}-1)2^{\frac{(\epsilon-1)}{2}k}}$. Exponential unbounded violation of order of $O\left((2^{\frac{1-2\epsilon - \sigma}{2}}\eta F)^{k}\right)$ is observed if $\epsilon+2\sigma < 2 \log_{2} (\eta F) + 1$ (similar analysis can be performed for $N=\lfloor{d^{1-\sigma}}\rfloor$ leading however, to more complicated formulas). It is remarkable that for any fidelity and efficiency satisfying $\eta F > \tfrac{1}{\sqrt{2}}$, there exist $\epsilon$ and $\sigma$ such that violation grows exponentially with the number of the entangled pairs $k$. 

\begin{figure}\centering
\includegraphics[height=6cm]{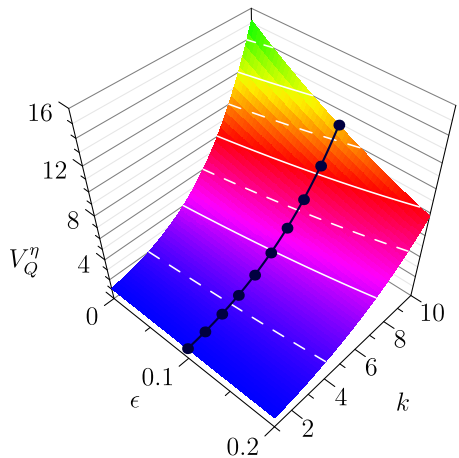}
\caption{Quantum violation of steering inequality ($\ref{quantum-violation}$) as a function of $k$ copies of a singlet state and the parameter $\epsilon$ setting the relaxed MUB condition. The best violation is obtained for $\epsilon =0$, i.e. the original MUBs. We took fidelity of a singlet state $F=0.98$, detection efficiency $\eta=0.95$ and $\sigma=0$. The dark blue line corresponds to the plot depicted in Fig.~\ref{fig:violation_eta} for $\eta=0.95$.}
\label{fig:violation_epsilon}
\end{figure}
\begin{figure}\centering
\includegraphics[height=5cm]{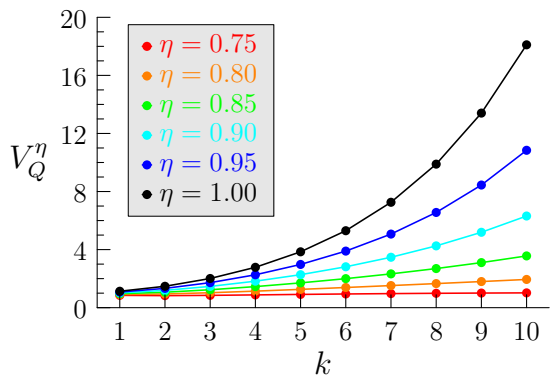}
\caption{Quantum violation $V_Q^{\eta}$ of steering inequality ($\ref{quantum-violation}$) computed for the relaxed MUB condition with $\epsilon=0.1$, singlet state fidelity $F=0.98$ and $\sigma=0$ as a function of detection efficiency $\eta$.}
\label{fig:violation_eta}
\end{figure}

Violation of the steering inequality ($\ref{quantum-violation}$) as a function of $k$ copies of a singlet state and the parameter $\epsilon$ setting the relaxed MUB condition is shown in Fig.~\ref{fig:violation_epsilon}. We assumed exemplary non-perfect fidelity of the singlet $F=0.98$, detection efficiency $\eta=0.95$ and $\sigma=0$. The best violation is obtained for the original MUBs, $\epsilon=0$. Relaxing this condition results in monotonic decrease of violation for increasing $\epsilon$ but, its exponential character is preserved. Fig.~\ref{fig:violation_eta} depicts dependence of violation on the efficiency $\eta$ for fidelity $F=0.98$, $\epsilon=0.1$ and $\sigma=0$. The minimal required efficiency to observe the violation equals $\eta=0.86$. Both figures reveal an interesting property of quantum steering: exponential decay of the global fidelity $F^{k}$ and efficiency $\eta^{k}$, observed for multi-pair sources of entangled qubits, is suppressed by exponential number of measurements. The latter is still possible to handle for small number of qubits (as it is done in tomographic experiments). In fact, this proposal has recently been demonstrated for the case of $k=4$ copies, and $N=2$ settings corresponding to two MUB elements~\cite{Li2016}. The main challenge in these types of experiments is extending the number of settings and generation of more MUBs. This seems to require a significant nonlinearity, a resource which is rare. Hence, below we shall consider a system where an unbounded violation of a steering inequality is possible using solely linear optics.

We will now employ a quantum-optical scheme based on a parametric-down-conversion source but working in a regime where it generates polarization entangled squeezed vacuum states. Their quantum correlations posses the same rotational invariance as a usual two-photon polarization singlet and can be seen as two copies of approximate original EPR correlations. Due to this property, these states have recently been successfully used to reveal a Bell nonlocality which does not vanish in the limit of large population~\cite{Rosolek2015}. Using the same key feature and implementing the generalized MUBs by merely polarization rotations, we will show that entangled squeezed vacuum states lead to unbounded violation of our steering inequality.

Entangled squeezed vacuum is a superposition of $2d$-photon polarization Bell-singlet states $\ket{\psi_{d}} = \frac{1}{\sqrt{d+1}}(a_{H}^{\dagger}b_{V}^{\dagger}-a_{V}^{\dagger}b_{H}^{\dagger})^{d} \ket{0}$ with a probability amplitude $\lambda_{d}$, $\ket{\Psi} = \sum_{d=0}^{\infty}\lambda_{d} \ket{\psi_{d}}$, where $a^{\dagger}$ ($b^{\dagger}$) is creation operator for a spatial mode $a$ ($b$) and $H$ ($V$) denotes horizontal (vertical) polarization. Perfect correlations present in each multi-particle polarization singlet are manifested by equal photon numbers in orthogonal polarizations in the spatial modes:
\begin{equation}
\ket{\psi_{d}} = \frac{1}{\sqrt{d+1}} \sum_{m=0}^{d} \left(-1\right)^{m}\ket{m_{H}, (d-m)_{V}}_{a} \ket{(d-m)_{H}, m_{V}}_{b}.
\label{Psi-N}
\end{equation}
They are preserved with respect to the global rotations of polarization. Each spatial mode in $\ket{\psi_{d}}$ contains fixed number of particles equal $d$. Projections on Fock states constitute a natural framework for Bob's photon counting measurements. They reveal not only the correlations but lead to post-selective creation of $\ket{\psi_{d}}$ from the squeezed vacuum state $\ket{\Psi}$. Thus, for simplicity, in our considerations we will now focus on a particular $|\psi_{d}\rangle$.

In order to show unbounded violation (\ref{quantum-violation}) for (\ref{Psi-N}) we will adopt a strategy similar to the one used in~\cite{Rosolek2015}: we will examine the correlations after applying incremental rotations on the spatial modes. In fact, due to rotational symmetry of $|\psi_{d}\rangle$, it is enough to consider the relative rotation angle between the modes $a$ and $b$, further denoted by $\theta_x$. Each mode represents a $(d+1)$-dimensional Hilbert space spanned by one of the basis chosen from a collection of $N$ bases, enumerated by $x=1, \dots, N$, generated by an appropriate polarization rotation:
\begin{equation}
\ket{\phi^{m}(\theta_x)} := \ket{(d-m)_{H+\theta_x}, m_{V+\theta_x}}.
\label{polarizationBases}
\end{equation}
They correspond to Bob's $\{\phi_{x}^{m}\}$ bases discussed in the Theorem and represent generalized MUBs (see Fig.~\ref{fig:suppl_overlap} in the Supplementary Information). We then have the following

\begingroup\it
Proposition. Given a set of $N$ Bob's measurement bases $\{ \lvert \phi^{m}_{x}\rangle \} := \{  \lvert \phi^{m}(\theta_{x})\rangle \}$ with $m=0, \ldots, d$ and $x=1, \ldots,N$, defined by some set of angles $0\leq \theta_{x} < \tfrac{\pi}{2}$, the maximal overlap $C=\max_{x,y,a,b}| \langle\phi_{x}^{a}\vert \phi_{y}^{b}\rangle|$ equals the  maximal overlap between $\{ \lvert\phi^{m}(0)\rangle \}$ and $\{ \lvert\phi^{m}(\theta)\rangle \}$ with $\theta=\min_{x,y} \bigl\lvert\theta_{x}  - \theta_{y}\bigr\rvert$:
\begin{align}
C(\theta,d)={}& \max_{m,n}\bigl\lvert \braket{\phi^{n}(0)\vert\phi^{m}(\theta)}\bigr\rvert =  \nonumber \\
{}={}&\sqrt{\binom{d}{q_{\theta,d}}}(\cos\theta)^{d} (\tan\theta)^{q_{\theta,d}}
\label{C}
\end{align}
where $q_{\theta,d}:=\lfloor d\sin^2\theta-\cos^2\theta\rfloor + 1$ and $\lfloor \dots \rfloor$ denotes the floor function. $C(\theta,d)$ goes to zero as fast as $1/\sqrt[4]{d}$.
\endgroup

Including experimental imperfections in their simplest form, we assume equal efficiency $\eta$ for each of the two detectors at Alice's. For the multi-particle Bell-singlet states (\ref{Psi-N}) this modifies the quantum value of the steering functional to $\eta^{d} S_{\text{Q}}$ and condition (\ref{quantum-violation}) to $V_{\text{Q}}^{\eta} \geq \dfrac{\eta^{d} N(d)}{1+ (N(d)-1) C(\theta,d)}$. Non-unit detection efficiency prevents from observation of the unbounded violation, which is expected. However, if the losses are moderate it will be possible to witness quantum steering for multi-particle quantum states. The price to pay with respect to our abstract result presented in the Theorem is that in presence of losses the number of settings grows with the local dimension quicker. Large number of measurements requires in turn very high accuracy in the polarization rotations.

\begin{figure}\centering
\includegraphics[width=\columnwidth]{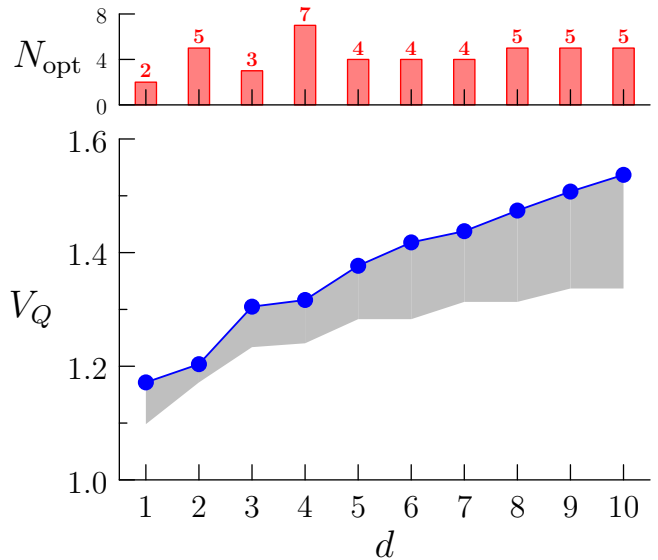}
\caption{Quantum violation $V_Q$ of steering inequality (\ref{quantum-violation}) for multi-particle Bell-singlet states $\lvert\psi_{d}\rangle$ (\ref{Psi-N}) and rotation angle $\theta=\tfrac{\pi}{2N}$. The upper inset shows the optimal number of settings $N_{\text{opt}}$ for a given $d$ used in this computation.  Grey area indicates a range of values of $V_Q$ for $N\not=N_{\text{opt}}$.}
\label{fig:violation}
\end{figure}

Fig.~\ref{fig:violation} depicts the violation $V_Q$ as a function of the local dimension $d$, for the optimal angle $\theta=\tfrac{\pi}{2N}$ and number of settings $N_{\text{opt}}$. The dependence $N_{\text{opt}}(d)$ shown in the upper inset of Fig.~\ref{fig:violation} is rather complex (see the Supplementary Information). Similarly, Fig.~\ref{fig:violation_eta2} depicts $V_Q^{\eta}$ as a function of the dimension $d$ and detection efficiency $\eta$.  As expected, the violation gets stronger for increasing population of the system, i.e. larger $d$-s, but also for higher detection efficiencies. However, we notice discrepancy between the value of critical efficiency for $d=1$ shown in this figure $\eta=0.86$, and the one reported in literature $\eta=0.62$~\cite{Smith2012}. We think that it reflects the fact that our estimation of the classical bound is not tight.

\begin{figure}\centering
\includegraphics[width=\columnwidth]{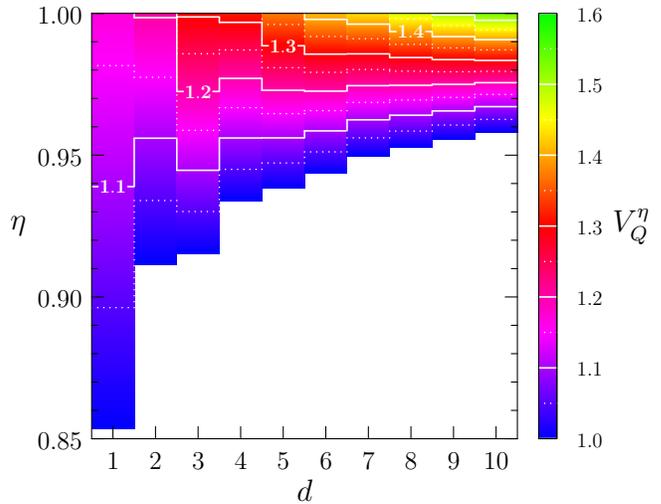}
\caption{Quantum violation $V_{\text{Q}}^{\eta}$ of steering inequality (\ref{quantum-violation}) computed for multi-particle Bell-singlet states $\lvert\psi_{d}\rangle$ (\ref{Psi-N}), detection efficiency $\eta$ and the best $N$ found for given $d$ (see Fig.~\ref{fig:violation}). White area indicates no violation.}
\label{fig:violation_eta2}
\end{figure}

{\it Discussion.} We would like to emphasize the following two aspects of the result we present. This is the first quantum steering inequality which is formulated in an error-tolerant way. Its robustness with respect to non-perfect measurement settings stems from its link to the entropic Deutsch--Maassen--Uffink uncertainty relation: the same parameter governs strength of both inequalities. Violation of this inequality with multi-particle quantum correlations seems feasible. We have suggested two experimental proposals. The first one is a proof of concept, which may serve as an illustrative example of unbounded violation of steering inequality involving relaxed MUB measurements. It can be performed with small number of two-photon singlet states and measurements. The second proposal seems particularly remarkable and stunning: this is a linear optics experiment which, also for the first time, can confirm that increasing arbitrarily slowly the local dimension of a maximally entangled state is sufficient to observe an unlimited discrepancy between the classical and quantum description of the experiment shown in Fig.~\ref{fig:scheme}.

Our result highlights very clear distinction between steering and Bell-nonlocality correlations. Violation of Bell inequalities with small number of settings vanishes for singlet states build from macroscopic qubits~\cite{Stobinska2012,Stobinska2014} or is constant for multi-particle Bell-singlet states~\cite{Rosolek2015}. In the case of quantum steering, unbounded violation is possible for multi-particle generalization of the original EPR correlations and almost constant number of settings in the form of MUBs, or their feasible counterparts. The latter applied to multiple copies of a singlet state may enable violation of order of $O(\sqrt{d})$ observed experimentally. On the other hand, multi-particle bipartite steering based on polarization entangled squeezed vacuum allows violation of order of $O(\sqrt[4]{d})$ at its best. Its realization necessitates linear optics equipped with photon-number-resolving detection and the generalized MUBs are implemented merely by polarization rotations, equally easy for every value of the local dimension. In our considerations we took into account experimental imperfections in the form of losses. Further analysis of the possibility of loophole-free violation and the role of fair sampling assumption similar to~\cite{Berry2010} remains to be addressed.

Our approach is very likely to be developed and exploited in several quantum information tasks which are of great technological interest. For example, recently the idea of partial characterization of  the devices has been developed and has been expressed in terms of semi-device-independent scenarios~\cite{Goh2015}. This approach lies between the two extremes of device-independent and tomographic strategies and  quantum steering is its natural prototype. Multi-particle bipartite steering can be used for randomness certification and may foster development of optimal strategies in quantum random access codes.

The above potential applications result also from the fact that our result is getting feasible although, technological requirements for an experimental setup are high. Polarization entangled squeezed vacuum states produced in parametric down conversion process, with mean photon number of the order of ten, are available in laboratories~\cite{Iskhakov2011,Iskhakov2012,Pan2012}. The states (\ref{Psi-N}) are created postselectively: each time Alice or Bob takes a measurement, the number of photons in each mode is known. Experiments could employ the techniques of~\cite{Radmark2009,Wieczorek2008} and integrated optics setups equipped with superconducting transition-edge sensors (TESs)~\cite{Lita2008}, which posses photon-counting efficiencies near $100\%$ and have well-resolved photon-number peaks, up to around ten photons~\cite{Gerrits2012}. For an alternative experimental implementation it is worth considering the angular momentum of light~\cite{Mair2001,Bobrov2015,Krenn2015}.

Our findings indicate new directions in the future research on quantum steering, by posing several open questions. The most intriguing is: How far does the general scaling in the number of settings $N$ go for the best steering monogamy relations? In this context the first results on that type of monogamy \cite{Milne2014} need further development. To address these issues it might be useful to analyze  the possible cases of steering inequalities with unequal number of settings at Alice's and Bob's. Existence of the monogamy relations here is conceivable if the number of settings for the state estimation at Bob's was of the order of the local dimension $d$. Then the asymmetric steering functional might satisfy a no-go limit in the form of an upper bound which value is fixed by diverging number of settings. Verifying this question is an important problem since it would determine what is the relation between Bell correlations and steering inequalities from the perspective of monogamy relation on the ground of no-signaling theories. Another interesting puzzle is: Can an asymmetric steering functional reveal quantumness in a more adequate way than the original one (cf.~\cite{Gheorghiu2015})? If yes, its potential applications in information processing may be worth examining. Our result highlights this point and we are planning to consider it in our future work.

Our findings open new conceptual and experimental perspective for steering inequalities and therefore, naturally stimulate questions in the context of quantum-information resources theory. For example, if and how the recent application of rigidity of quantum steering to device-independent one-sided computation~\cite{Gheorghiu2015} can be adopted and made useful in our case? Further, it would be interesting to examine our result on the ground of general probabilistic theories referred to the receiver's (Bob) system (cf.~\cite{Massar2015} and references therein).

We hope that the results and discussion presented in this paper will stimulate the future research on quantum correlations in a broader context, including their applications and demonstrations.

{\it Methods.} The proof of our Theorem involves a surprisingly powerful modification of the former analysis presented in~\cite{Marciniak2015}: it builds-in tolerance for possible errors in measurement settings. This was not possible in the framework of~\cite{Marciniak2015}. In ~\cite{Marciniak2015} the steering functional $S_{\text{LHS}}$ is bounded by the operator norm of a Gramm matrix $G$ with matrix elements equal $G_{ax,by}:=\sqrt{p(a|x)\,p(b|y) }\langle\phi_{x}^{a}\lvert\phi_{y}^{b}\rangle$. Its estimation is very troublesome because the matrix involves probabilities of an unknown value, resulting from the local hidden state model. Hence, any upper bound on the matrix norm $\lVert G\rVert_{\infty}=\sup_{\lvert v\rangle, \lVert v\rVert=1} \lVert Gv\rVert$ must optimize over all probability values. Our technique was to exploit the tensor structure appearing due to double indices in a new, much efficient way. This was done by means of specific decomposition of the $x,y$ part of the structure, which employs mutually orthogonal shift $n \times n$ matrices. This key trick allows to go much farther beyond the rough estimates of hidden probabilities parameters used so far. Indeed, the latter allows to bound the maximal eigenvalue of the product $G^{\dagger}G$ thus, the singular value of $G$ as well, by factorizing out the probabilities due to their normalization rather then simply binding them all by unity. This allows to control the degree of mutual unbiasedness among all the bases individually i.e. from one pair to another in terms of Deutsch--Maassen--Uffink parameters.

Derivation of the Proposition exploited the properties of rotational symmetry of multi-particle Bell-singlet states and the fact that the overlaps between multi-particle polarization bases (\ref{polarizationBases}) generate a probability distribution
\begin{equation}
p(m,n|\theta)=\bigl\lvert \braket{\phi^{n}(0)\vert\phi^{m}(\theta)}\bigr\rvert^2,
\end{equation}
which corresponds to a two-dimensional continuous-time quantum walk~\cite{Konno2004,Watanabe2008} (see the Supplementary Information). Thus, $p(m,n|\theta)$ does not follow the Central Limit Theorem and its maximal values are taken at the boundaries of its domain. Further, to obtain the formula (\ref{C}), some specific properties of binomial coefficients were used~\cite{Pahor2006}. The asymptotic value of the maximal overlap was computed using the Stirling's formula. Optimal number of settings $N_{\text{opt}}$ and set of angles for given $d$ and $\theta$ were found by a numerical algorithm. Technical details are presented in the Supplementary Information.

Finally, we would like to stress that the experimental implementation requires correct space-time arrangement~\cite{Wittmann2012} and any faster-than-light signaling due to apparatus correlations must be excluded (in order to avoid controversies similar to the ones which arose in some tests of Bell inequalities).

\medskip

\begin{acknowledgments}
MS and AB were supported by the EU 7FP Marie Curie Career Integration Grant No. 322150 ``QCAT'', NCN grant No. 2012/04/M/ST2/00789, MNiSW co-financed international project No. 2586/7.PR/2012/2 and MNiSW Iuventus Plus project No. IP 2014 044873. AR thanks for support ERC AdG QOLAPS and NCN grant No. 2014/14/M/ST2/00818. PH acknowledges support of ERC AdG QOLAPS.  The authors thank Ryszard Horodecki for a discussion.
\end{acknowledgments}

\section*{Contributions}

MS \& PH supervised the project, suggested the idea and its physical implementation. They provided the discussion of results and wrote up the paper. AR participated in discussions, suggested and proved the technical version of the theorem. AB developed numerical methods and analytical computations supporting the proposition in discussion with MS. AB prepared the figures.

%%%%%%%%%%%%%%%%%% Supplementary Information %%%%%%%%%%%%%%%%%%%%%%%%%%%

\onecolumngrid

\renewcommand{\theequation}{SI.\arabic{equation}}
\setcounter{equation}{0}

\clearpage

\begin{center}
{\Large\bf Supplementary Information:\\\strut Quantum steering inequality with tolerance\\\strut for measurement-setting-errors:\\\strut
experimentally feasible signature of unbounded violation}
\end{center}

\begin{quote}
The Supplementary Information comprises detailed proofs of the Theorem and the Proposition from the main text, as well as an elementary argument that any unbounded violation of Bell inequalities requires divergence of at least one party's settings.
\end{quote}

\section*{Main Theorem}

\noindent
\begingroup\it
Given a quantum steering scenario involving $x\in\{1,\dots,N\}$ settings, $a \in\{1,\dots,d\}$ outcomes, and the set of orthonormal eingenbases $\left|\phi_{x}^{a}\right\rangle $ defining the receiver's (Bob's) measurements, the LHS steering functional is bounded from above
\begin{equation}
S_{\text{LHS}} \leq 1+\sum_{i=1}^{N-1} C_{i},
\end{equation}
where $C_i=\max_{x }C_{x \,N+x-i}$ and and
where  $C_{xy}:=\max_{a,b} |\braket{\phi_{x}^{a}\vert\phi_{y}^{b}}|$
for $x,y \in { 1, \ldots, N}$  as in the Deutsch--Maassen--Uffink uncertainty
relations. This implies:
\begin{equation}
V_{\text{Q}} \geq \dfrac{N}{1+\sum_{i=1}^{N-1} C_{i}},
\label{supeq:V_Q}
\end{equation}
From the above in particular the weaker variant follows, namely
\begin{equation}
V_{\text{Q}} \geq \dfrac{N}{1+ (N-1) C},
\end{equation}
where $C=\max_{i} C_{i}=\max_{x\neq y}C_{x,y}$.
\endgroup

\subsection*{Proof}

First, we will compute the quantum value $S_{Q}(F)$ for the steering functional $F$ defined as a set $F=\left\{ \left|\phi_{x}^{a}\left\rangle \right\langle \phi_{x}^{a}\right|:\: a=1,\ldots,d,\, x=1,\ldots,N\right\} $. In our case it is enough to show that $S_{Q}\bigl(F\bigr) \leq N$, where $N$ is the number of bases
\begin{align}
S_{Q}\bigl(F\bigr)
=
\mathrm{Tr}\left\{
  \sum_{x=1}^N \sum_{a=1}^{d}
  \ket{\phi_{x}^{a}} \bra{\phi_{x}^{a}} \sigma_{x}^{a}\right\}
\leq{}&
\mathrm{Tr}\left\{
  \sum_{x=1}^{N} \sum_{a=1}^{d}
  \ket{\phi_{x}^{a}} \bra{\phi_{x}^{a}}
  \sum_{a'}\sigma_{x}^{a'}\right\}\nonumber\\
{}={}&
\mathrm{Tr}\left\{
  \sum_{x=1}^{N} \sum_{a=1}^{d}
  \ket{\phi_{x}^{a}} \bra{\phi_{x}^{a}} \rho_{x} \right\}\\
{}={}&
  \sum_{x=1}^{N} \underbrace{\sum_{a=1}^{d} p_{x}(a|x)}_{=1}=N\nonumber\\
{}\Downarrow{}&\nonumber\\
S_{Q}\leq{}& N.
\end{align}

Second, we will describe a general method of computation of the classical bound $S_{\text{LHS}}\bigl(\ket{\phi_{x}^{a}}\bra{\phi_{x}^{a}}\bigr)$. Let $\sigma_{x}^{a}=\sum_{\lambda}q_{\lambda}\,p_{\lambda}(x|a)\,\sigma_{\lambda}$. Then, the following inequality holds
\begin{align}
\mathrm{Tr}\left\{
  \sum_{x=1}^{N}\sum_{a=1}^{d}
  \ket{\phi_{x}^{a}} \bra{\phi_{x}^{a}} \sigma_{x}^{a}\right\}
={}&
\sum_{x=1}^{N}\sum_{a=1}^{d}
  \mathrm{Tr}\left\{
  \ket{\phi_{x}^{a}}\bra{\phi_{x}^{a}}
  \sum_{\lambda} q_{\lambda}\,p_{\lambda}(a|x)\,\sigma_{\lambda}\right\}\nonumber\\
{}={}&
\sum_{\lambda} q_{\lambda}
\sum_{x=1}^{N}\sum_{a=1}^{d}
  \mathrm{Tr}\bigl\{
    \ket{\phi_{x}^{a}} \bra{\phi_{x}^{a}}
    p_{\lambda}(a|x)\,\sigma_{\lambda}\bigr\}\nonumber\\
{}\leq{}&
\sup_{\lambda}\left\lVert
  \sum_{x=1}^{N}\sum_{a=1}^{d}
  \ket{\psi_{x,\lambda}^{a}} \bra{\psi_{x,\lambda}^{a}}\right\rVert,
\end{align}
where $\ket{\psi_{x,\lambda}^{a}}=\sqrt{p_{\lambda}(a|x)}\ket{\phi_{x}^{a}}$. For any $\lambda$, let $G_{\lambda}=\sum_{x,y=1}^{N}\sum_{a,b=1}^{d}\braket{\psi_{x,\lambda}^{a}\vert\psi_{y,\lambda}^{b}} \ket{x} \bra{y} \otimes \ket{a} \bra{b}$. Using the purification of $\sum_{x=1}^{N}\sum_{a=1}^{d}\ket{\psi_{x,\lambda}^{a}} \bra{\psi_{x,\lambda}^{a}}$ and its Schmidt decomposition, we can show that
\begin{equation}
\left\lVert \sum_{x=1}^{N}\sum_{a=1}^{d} \ket{\psi_{x,\lambda}^{a}} \bra{\psi_{x,\lambda}^{a}} \right\rVert = \left\lVert G_{\lambda}\right\rVert.
\end{equation}
In further considerations we will omit the index $\lambda$. Let us define the shift operator $S:\mathbb{C}^{N}\rightarrow\mathbb{C}^{N}$ acts on the bases vectors in the following way:
\begin{equation}
S\left|k\right\rangle =\left|k+1\right\rangle \quad\text{{mod}}\, N,
\end{equation}
and observe that $\sum_{i=1}^{N} S^{i}=\mathbb{I}$ -- every element of $\mathbb{I}$ is equal to $1$.
 We decompose $G$ in the following way
\begin{equation}
G=\sum_{x,y=1}^{N}\ket{x} \bra{y} \otimes G_{xy} = \sum_{i=1}^{N} A_{i},
\end{equation}
where
\begin{equation}
A_{i} = \sum_{(x,y)\in\mathcal{S}_{i}} \ket{x} \bra{y} \otimes G_{xy},
\end{equation}
and the set $\mathcal{S}_{i}=\bigl\{(x,y): S_{xy}^{i}=1\bigr\}$. Next, we use the following fact
\begin{equation}
\left\Vert G\right\Vert \leq\sum_{i=1}^{N}\left\Vert A_{i}\right\Vert .
\end{equation}
Hence, in order to estimate the norm of $G$ we have to estimate the norm of $A_{i}$. This is just the maximal singular value of $A_{i}$ or, equivalently, the maximal eigenvalue of $A_{i}A_{i}^{\dagger}$, squared. Since this operator is block diagonal ($S^{i}$ are permutation operators), we have to calculate the maximal singular value of $G_{xy}$, taking into account the proper index of $i$. To this end, let us estimate this singular value of $G_{xy}$, which possess the following general form
\begin{equation}
G_{xy}  =\sum_{a,b=1}^{d}\alpha_{xy}^{ab} e^{i\psi_{xy}^{ab}}\sqrt{p(a|x)\,p(b|y)} \ket{a} \bra{b},
\end{equation}
where $\alpha_{xy}^{ab}e^{i\psi_{xy}^{ab}}=\left\langle \phi_{x}^{a}|\phi_{y}^{b}\right\rangle $
and $\alpha_{xy}^{ab}=\left|\left\langle \phi_{x}^{a}|\phi_{y}^{b}\right\rangle \right|$
while $\psi_{xy}^{ab}$ are phases for given indices $a$, $b$, $x$ and $y$. This results in
\begin{align}
G_{xy}G_{xy}^{\dagger} ={}&
\sum_{a,b,a',b'=1}^{d} \alpha_{xy}^{ab}\alpha_{xy}^{a'b'}
  e^{i\bigl(\psi_{xy}^{ab}-\psi_{xy}^{a'b'}\bigr)}
  \sqrt{p(a|x)\,p(b|y)\,p(a'|x)\,p(b'|y)}
  \ket{a}\bra{b}\ket{b'}\bra{a'}\nonumber\\
{}={}&
\sum_{a,b,a'=1}^{d}\alpha_{xy}^{ab}\alpha_{xy}^{a'b}
  e^{i\bigl(\psi_{xy}^{ab}-\psi_{xy}^{a'b}\bigr)}
  p(b|y) \sqrt{p(a|x)\,p(a'|x)}
  \ket{a}\bra{a'}.
\label{eq:supplmat_re}
\end{align}
Here we use the fact that $G_{xy}G_{xy}^{\dagger}\geq 0$. This means that $\mathrm{Tr}\bigl\{G_{xy}G_{xy}^{\dagger}\bigr\}=\sum_{j=1}^{d}\lambda_{xy}^{j}$, where $\lambda_{xy}^{j}$ are eigenvalues of $G_{xy}G_{xy}^{\dagger}$. Let us denote the maximal eigenvalue as $\lambda_{xy}^{\text{max}}=\max_{j}\bigl\{\lambda_{xy}^{j}\bigr\}$. From (\ref{eq:supplmat_re}) we obtain the maximal singular value of $G_{xy}$, $\sigma_{xy}^{\text{max}}$
\begin{align}
\bigl(\sigma_{xy}^{\text{max}}\bigr)^{2}=\lambda_{xy}^{\text{max}}\leq{}
\sum_{i=1}^{d}\lambda_{xy}^{i}={}&\mathrm{Tr}\Bigl\{G_{xy}G_{xy}^{\dagger}\Bigr\}
\nonumber\\
{}={}&
\mathrm{Tr}\left\{
  \sum_{a,b,a'=1}^{d}\alpha_{xy}^{ab}\alpha_{xy}^{a'b}e^{i\bigl(\psi_{xy}^{ab}-\psi_{xy}^{a'b}\bigr)} p_{y}^{b} \sqrt{p_{x}^{a} p_{x}^{a'}} \ket{a} \bra{a'}\right\}\nonumber\\
{}={}&
\sum_{a,b=1}^{d}\bigl(\alpha_{xy}^{ab}\bigr)^{2} p_{y}^{b} p_{x}^{a}
\leq
\sum_{a,b=1}^{d}\bigl(\alpha_{xy}^{max}\bigr)^{2}p_{y}^{b}p_{x}^{a} = \bigl(\alpha_{xy}^{\text{max}}\bigr)^{2}.
\end{align}
Therefore, in order to estimate the norm of $G$ we must calculate the maximal absolute value of the overlap between vectors $\alpha_{xy}^{\text{max}}$ of the basis given by the number $x,y$. Then, the norm
\begin{equation}
\lVert G\rVert \leq \sum_{i=1}^{N} C_{i},
\end{equation}

where $C_i=\alpha_{i}^{\text{max}}=\max_{x,y} \bigl\{\alpha_{xy}^{max}: (x,y)\in\mathcal{S}_{i}\bigr\}$ ($C_i$ is just Mussen-Uffink value for each $i$ ). Let us observe that for $x=y$  it is just identity  transformation between these two bases
(it corresponds to the case $i=N$) hence $\alpha_{N}^{\max}=1$ and

\begin{equation}
S_{\text{LHS}} \leq \lVert G\rVert \leq \sum_{i=1}^{N}  C_{i}\leq 1+\sum_{i=1}^{N-1} C_{i}.
\end{equation}
Finally, the violation of the steering inequality
\begin{equation}
V_{\text{Q}} \geq \dfrac{N}{1+\sum_{i=1}^{N-1} C_{i}},
\end{equation}
Which ends the proof of main theorem.
In the main text we assume the worst case to calculate the norm of $G$

\begin{equation}
S_{\text{LHS}} \leq \lVert G\rVert \leq1+ \sum_{i=1}^{N-1} C_{i}\leq 1+\left(N-1\right)\,C,
\end{equation}
where $C=\alpha^{\text{max}}= \max_{x\neq y} \bigl\{\alpha_{xy}^{\text{max}}\bigr\}=\max_{i}\bigl\{\alpha_{i}^{\text{max}}\bigr\} $ for  $i\neq N$. Thus,
\begin{equation}
V_{\text{Q}} \geq \dfrac{N}{1+\left( N-1\right)\,C}.
\end{equation}

At the end of our consideration let us define the set of matrices $O_{xy}=\left[\left\langle \phi_{x}^{a}|\phi_{y}^{b}\right\rangle \right]_{a,b=1}^{d}$ and consider the new matrix which has the form
\begin{equation}
O={\displaystyle \sum_{x,y=1}^{N}\left|x\left\rangle \right\langle y\right|\otimes O_{xy}}.
\end{equation}
Let us notice that matrix $O$ is just matrix $G$ without probabilities $p(a|x)$ and $p(b|y)$,  then we can formulate the main theorem in the stronger form (in the sense of the shorter summation):

Observation: If for the   matrix $G$ the matrix $O$ has a block Toeplitz form namely
$O_{xy}=O_{x+1,y+1}$ for any $x,y \in \{1,\ldots,N\}$  then the inequality for $ S_{\text{HLS}}$ takes the form
\begin{equation}
S_{\text{LHS}}\leq1+{\displaystyle \left(\left\lceil N'\right\rceil -\left\lfloor N'\right\rfloor \right)C_{\left\lfloor N'\right\rfloor +1}+2\sum_{i=1}^{\left\lfloor N'\right\rfloor }}C_{i}
\end{equation}
where $N'=\frac{N-1}{2}$ and $\lceil\ldots\rceil$, $\lfloor\ldots\rfloor$ are ceiling and floor functions respectively. Here we provide an explanation of this observation instead of the formal proof. This observation comes from symmetry of the  block Toeplitz Hermitian matrix $O$ what implies  $\alpha_{i}^{\text{max}}=\alpha_{N
-i}^{\text{max}}$ for $i=1,\ldots,\left\lfloor \frac{N-1}{2}\right\rfloor $.
%In fact the parameters
%\begin{equation}
 %\alpha_{i^{\text{max}}}=max_{xy}\sigma_{xy}^{\text{max}=
 %max_{x} \sigma(G_{x (N+x-i)mod N})
%\label{alfa}
%\end{equation}
 %where $\sigma$ is here the matrix norm.
Let us explain in detail what we mean by symmetry in this case. The reader should  notice that we have  $ N-1$ sets ${\cal O}_{i}=\{ O_{x (N+x- i)\, \text{mod}\, N} \}$
and the fundamental fact is that due to Toeplitz form each of the sets
is formed of the two matrices and  each of them are repeated
some number of times,  we may, slightly abusing the notation, write
${\cal O}_{i}=\{ O_i,  O_{N- i}\} $ for the Toeplitz matrices.
Now, due to the hermicity of the global matrix $O$ which determines that
$O_{xy}=O_{yx}^{\dagger}$ and only the sets   ${\cal O}_{i}$  for
$i=1,\ldots,\left\lfloor \frac{N-1}{2}\right\rfloor$ (if $N$ is odd otherwise we have to add one additional term in  summation, namely $ \alpha_{\left\lfloor \frac{N-1}{2}\right\rfloor+1}^{\max}$ ) should be taken into account, while
calculating the quantities $\alpha_i^{\max}$ since
any set for higher $i$ reproduces one of the previous ones modulo Hermitian
conjugate of its elements, which does not change the singular values of
the matrices contained in the set.

\section*{Analysis of non-signaling principle}

In this section we recall an elementary argument that any unbounded violation of Bell inequalities requires divergence of at least one party's settings (here this is Bob, but considering Alice makes no difference).

Quantum monogamy relation says that if we consider a no-signaling box $P(a,b_{1},\ldots,b_{n}|x,y_{1},\ldots,y_{n})$
among $n+1$ parties $A,B_{1},\ldots,B_{n}$, where each Bob has $n$ observables at his disposal, the following monogamy relation holds for Bell functionals $\mathcal{B}(A,B)$ \cite{Pawlowski2009}:
\begin{equation}
\sum_{i=1}^{n} \mathcal{B}(A,B^{(i)}) \leq n\, R_{\text{LHS}},
\end{equation}
with the original bipartite Bell inequality for Alice and Bob defined as $\mathcal{B}(A,B) \leq R_{\text{LHS}}$.
Let us assume that the Bell functionals are non-negative. Then, taking an arbitrary extension of an arbitrary no-signaling bipartite box $P_{\text{NS}}(a,b|x,y)$, $P_{\text{extension}}(a,b_{1},\ldots,b_{n}|x,y_{1},\ldots,y_{n})$ (in a sense that averaging  $P_{\text{extension}}$ over the $n-1$ pairs of variables $\{ b_{2},\ldots,b_{n};y_{2},\ldots,y_{n}\}$ reproduces $P_{\text{NS}}$), gives
\begin{equation}
\mathcal{B}_{P_{\text{NS}}}(A,B):= \mathcal{B}(A,B^{(1)}) \leq n\,R_{\text{LHS}}  - \sum_{i=2}^{n} \mathcal{B}(A,B^{(i)}) \leq  n\,R_{\text{LHS}},
\end{equation}
due to non-negativity of the functional. Note that minimizing over all the extensions within a given sub-theory of a no-signaling theory would give a bound within that ``subtheory'' (including quantum mechanics). Thus, for any no-signaling bipartite $P_{\text{NS}}(a,b|x,y)$, one has the Bell-violation rate bounded by $V_{Bell}(NS) = \frac{\mathcal{B}_{P_{\text{NS}}}(A,B) }{R_{\text{LHS}}} \leq n$. Since any bipartite quantum statistics represents a no-signaling box, it must obey the same bound.

\section*{The Proposition}

\noindent
\begingroup\it
Given a set of $N$ Bob's measurement bases $\{ \lvert \phi^{m}_{x}\rangle \} := \{ \lvert \phi^{m}(\theta_{x})\rangle \}$ with $m=0, \ldots, d$ and $x=1, \ldots,N$, defined by some set of angles $0\leq \theta_{x} < \tfrac{\pi}{2}$, the maximal overlap $C=\max_{x,y,a,b}| \langle\phi_{x}^{a}\vert \phi_{y}^{b}\rangle|$ equals the maximal overlap between $\{ \lvert\phi^{m}(0)\rangle \}$ and $\{ \lvert\phi^{m}(\theta)\rangle \}$ with $\theta=\min_{x,y} \bigl\lvert\theta_{x}  - \theta_{y}\bigr\rvert$:
\begin{equation}
C(\theta,d)={} \max_{m,n}\bigl\lvert \langle \phi^{n}(0)\vert \phi^{m}(\theta) \rangle\bigr\rvert = \sqrt{\binom{d}{q_{\theta,d}}}(\cos\theta)^{d} (\tan\theta)^{q_{\theta,d}},
\end{equation}
where $q_{\theta,d}:=\lfloor d\sin^2\theta-\cos^2\theta\rfloor + 1$% and $\lfloor \dots \rfloor$ denotes the floor function.
$C(\theta,d)$ goes to zero as fast as $1/\sqrt[4]{d}$.
\endgroup

\subsection*{Proof}

\noindent
Let us consider a two-mode state expressed in $\{H,V\}$ basis, which is rotated by the angle $\theta$
\begin{equation}
\begin{pmatrix}
b^\dagger_{H+\theta}\\b^\dagger_{V+\theta}
\end{pmatrix}
=
\begin{pmatrix}
\cos\theta& \sin\theta\\ -\sin\theta& \cos\theta
\end{pmatrix}
\cdot
\begin{pmatrix}
b^\dagger_H\\b^\dagger_V
\end{pmatrix}.
\end{equation}
Then, the state $\ket{d-m,m}$ in $\{H+\theta,V+\theta\}$ basis takes the following form in $\{H,V\}$
\begin{equation}
\ket{(d-m)_{H+\theta},(m)_{V+\theta}}
=
\begin{aligned}[t]
\dfrac{1}{\sqrt{(d-m)!\,m!}}
\sum_{p=0}^{d-m}
\sum_{q=0}^m
&\binom{d-m}{p}\binom{m}{q}(-1)^q (\cos\theta)^{m+p-q} (\sin\theta)^{d-m-p+q}\times{}\\
{}\times{}&\sqrt{(p+q)!(d-p-q)!} \ket{(p+q)_H, (d-p-q)_V}.
\end{aligned}
\end{equation}
Overlap with $\ket{(d-n)_H,(n)_V}$ equals to
\begin{equation}
O_d^{n,m}(\theta) = \sqrt{\dfrac{(d-n)!\,n!}{(d-m)!\,m!}}
\kern0.75em
\sideset{}{_2}\sum_{k=\max\bigl[-(d-n),(d-n)-2m\bigr]}^{\min\bigl[(d-n),2(d-m)-(d-n)\bigr]}
\kern0.5em
\binom{d-m}{\tfrac{d-n+k}{2}}\binom{m}{\tfrac{d-n-k}{2}}(-1)^{\tfrac{d-n-k}{2}} (\cos\theta)^{m+k} (\sin\theta)^{d-m-k},
\end{equation}
where $\sideset{}{_2}\sum f(d)$ symbol denotes sum of $f(d)$ for index $d$ increasing by 2, i.e.
\begin{displaymath}
\sideset{}{_2}\sum_{d=d_\text{min}}^{d_\text{max}} f(d) = f(d_\text{min}) + f(d_\text{min}+2) + \ldots + f(d_\text{max}-2) + f(d_\text{max}).
\end{displaymath}

\begin{figure*}[t]
\begin{tabular}{cccc}
& $\theta=\dfrac{\pi}{8}$& $\theta=\dfrac{\pi}{6}$& $\theta=\dfrac{\pi}{4}$\\
\raisebox{3cm}{a)}&
\includegraphics[width=5cm]{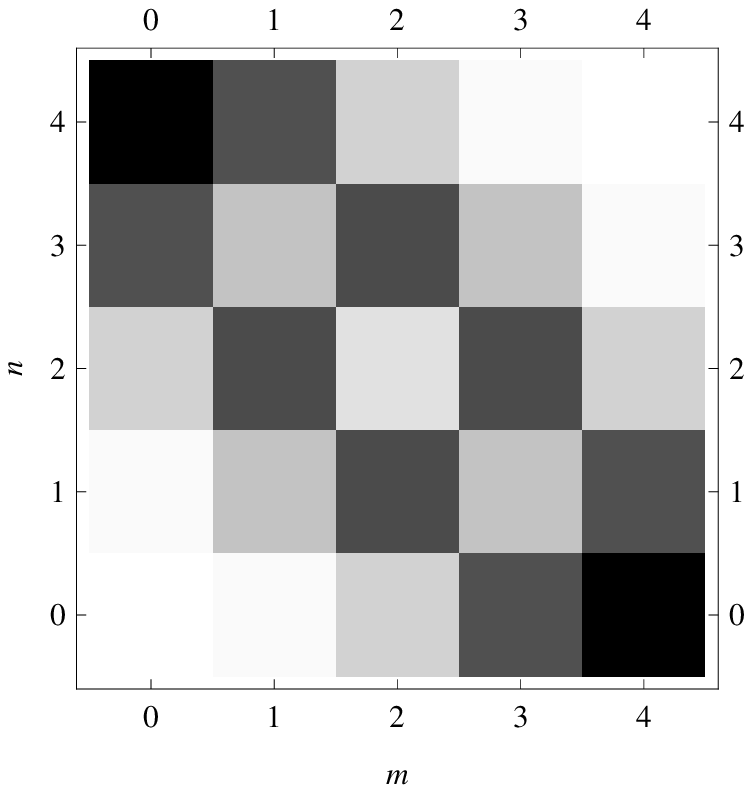}&
\includegraphics[width=5cm]{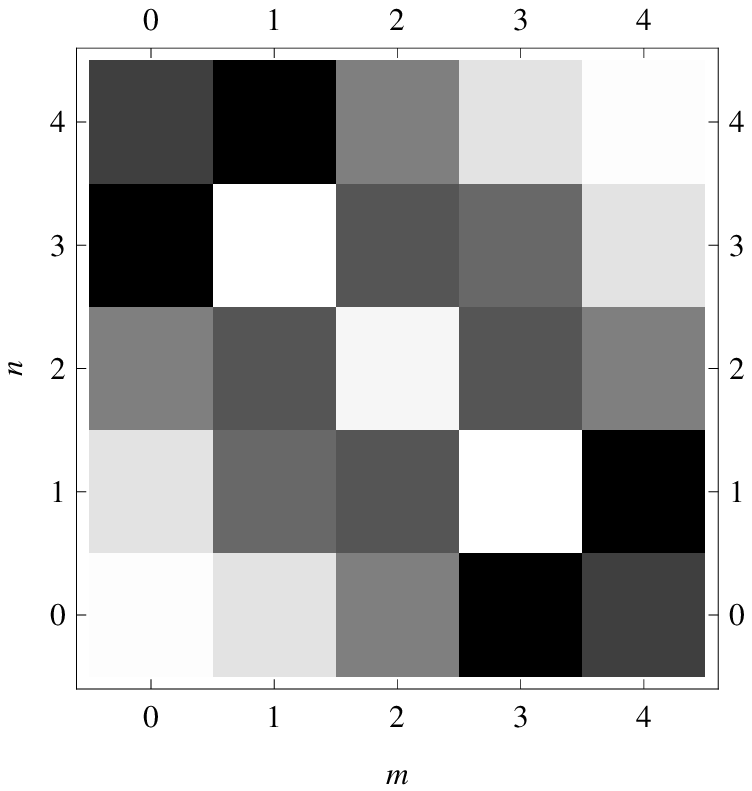}&
\includegraphics[width=5cm]{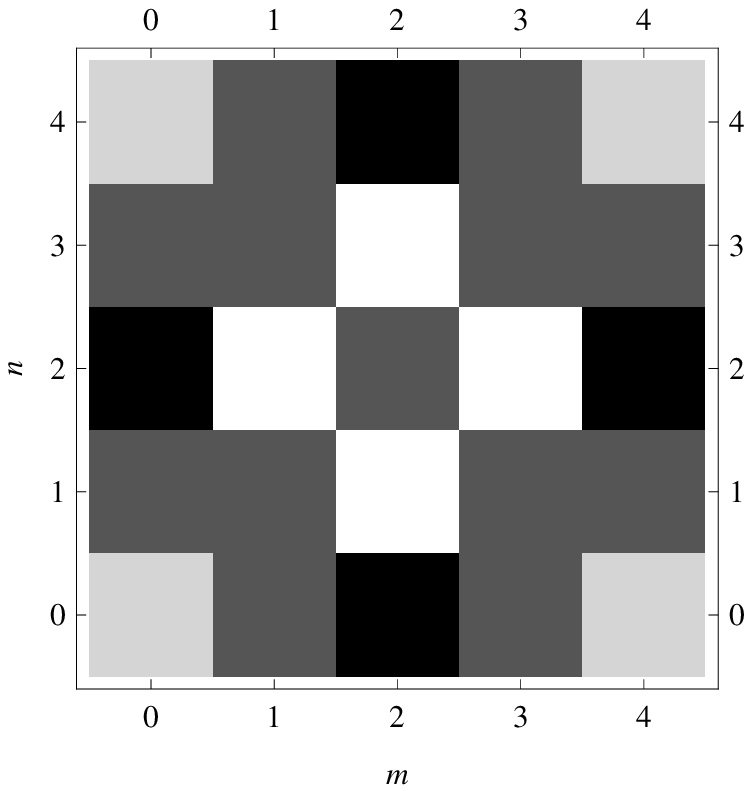}
\\
\raisebox{3cm}{b)}&
\includegraphics[width=5cm]{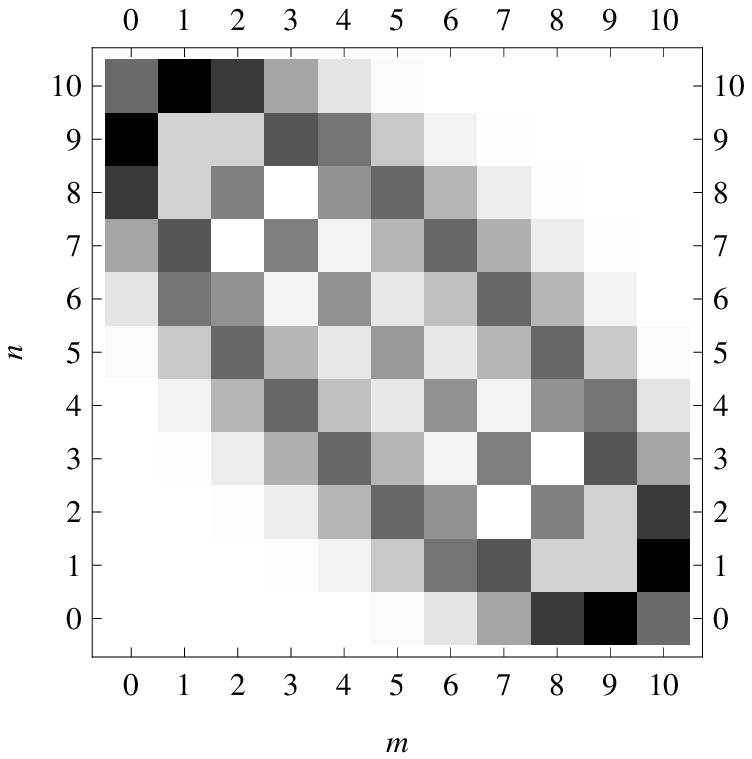}&
\includegraphics[width=5cm]{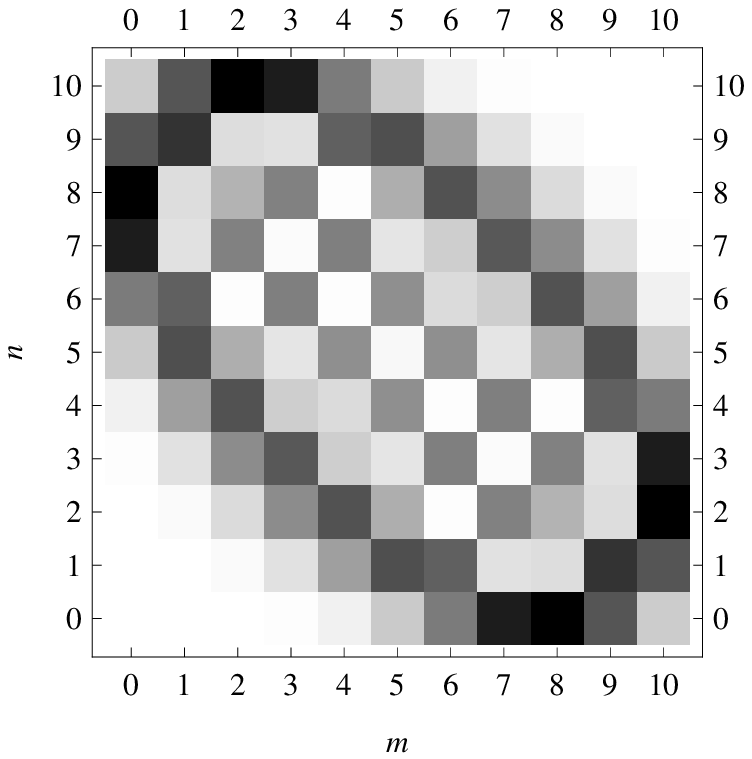}&
\includegraphics[width=5cm]{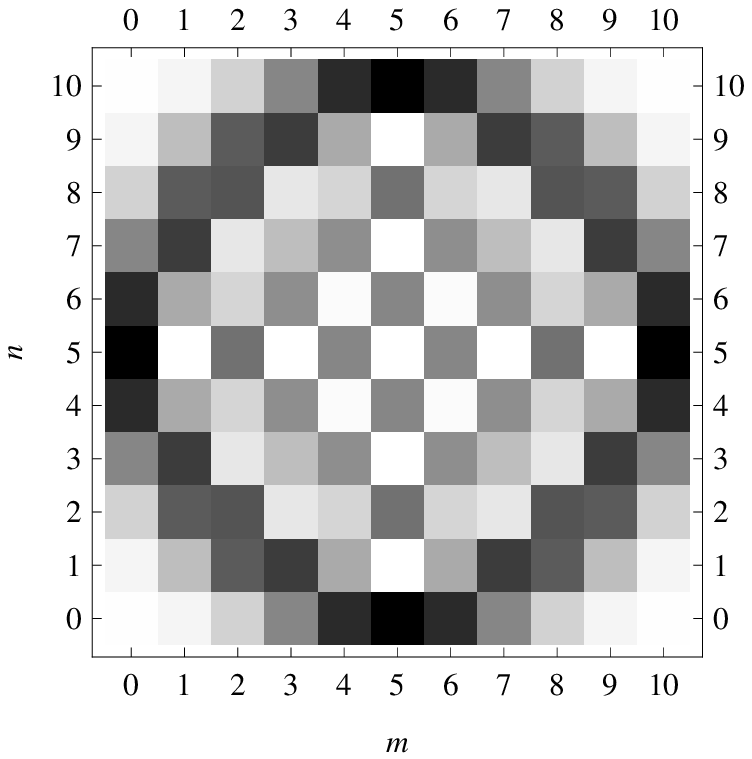}
\\
\raisebox{3cm}{c)}&
\includegraphics[width=5cm]{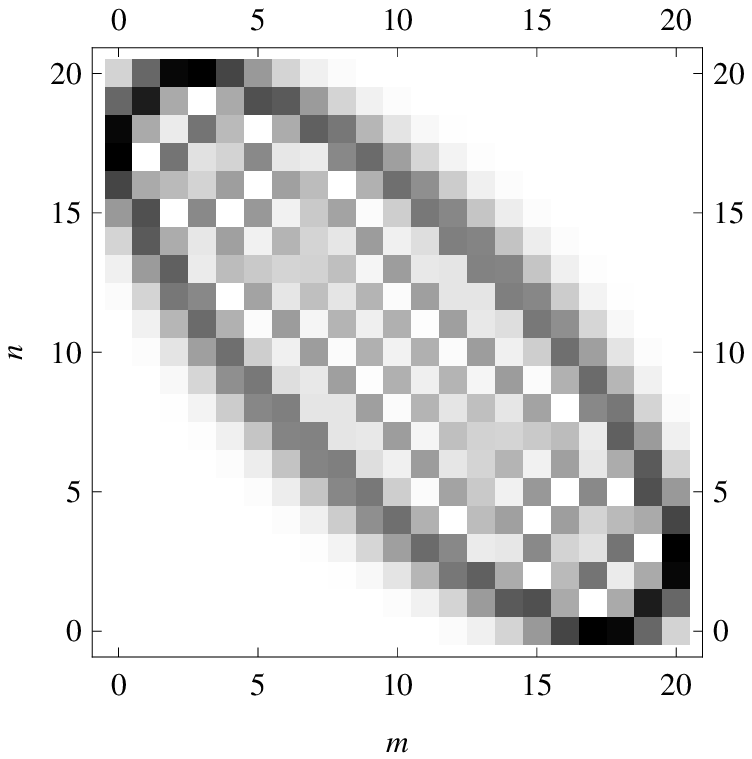}&
\includegraphics[width=5cm]{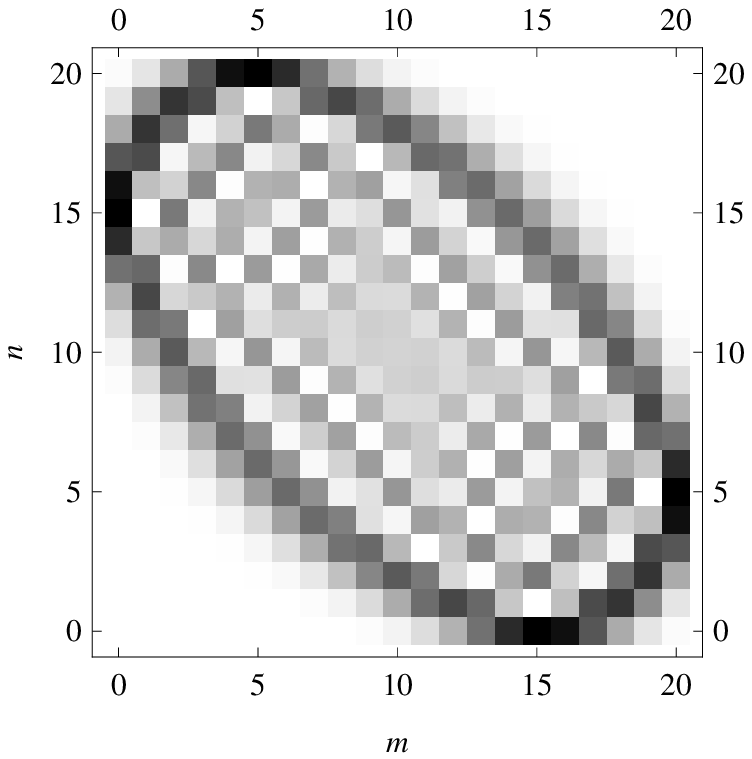}&
\includegraphics[width=5cm]{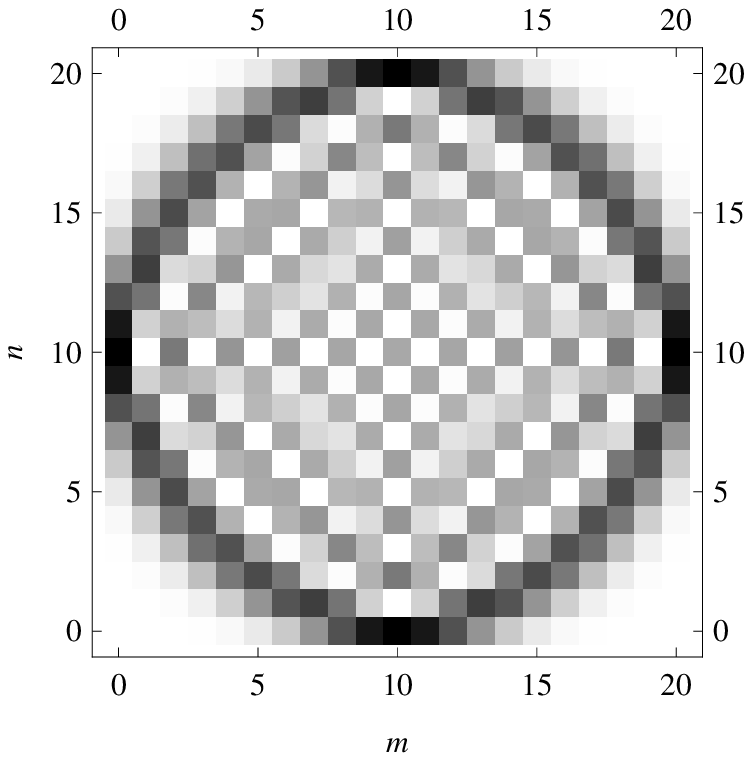}
\end{tabular}
\caption{$\lvert O_d^{n,m}(\theta)\rvert^2$ computed for a) $d=4$ b) $d=10$ and $d=20$ for $\theta=\pi/8$, $\theta=\pi/6$ and $\theta=\pi/4$. The maximal values of $\lvert O_N^{n,m}(\theta)\rvert^2$ always lie at $n\in\{0,d\}$ or $m\in\{0,d\}$.}
\label{fig:suppl_overlap}
\end{figure*}

\noindent
Fig.~\ref{fig:suppl_overlap} depicts $\lvert O_d^{n,m}(\theta)\rvert^2$ for $d=4,10,20$ and several values of $\theta$.
Numerics indicates that $\lvert O_d^{n,m}(\theta)\rvert^2$ reaches its all maximal values only for $n\in\{0,d\}$ or $m\in\{0,d\}$ and they are equal. Therefore, we focus exclusively at $n=0$ and $m=0$.  For $m=0$:
\begin{equation}
O_d^{n,0}(\theta) = \sqrt{\binom{d}{n}} (\cos\theta)^{d-n} (\sin\theta)^n,
\end{equation}
and for $n=0$:
\begin{equation}
O_d^{0,m}(\theta) = \sqrt{\binom{d}{m}} (-1)^m (\cos\theta)^{d-m} (\sin\theta)^m.
\end{equation}
In order to find maximal values of $\bigl\lvert O_d^{n,0}(\theta)\bigr\rvert^2$ and $\bigl\lvert O_d^{0,m}(\theta)\bigr\rvert^2$ one should note that they are just binomial coefficients of the form $\binom{d}{k} r^{d-k} (1-r)^k$ thus, they reach maximal values for $k_\text{max}=\lfloor d(1-r)-r\rfloor+1$ (see e.g.~\cite{Pahor2006}).  Thus, for $m=0$
\begin{equation}
n_{\text{max}} = \lfloor d\sin^2\theta-\cos^2\theta\rfloor+1,
\end{equation}
and for $n=0$
\begin{equation}
m_{\text{max}} = \lfloor d\sin^2\theta-\cos^2\theta\rfloor+1,
\end{equation}
which gives also $q_{\theta,d}=n_\text{max}=m_\text{max}$ for given $d$ and $\theta$. Finally, let us notice that
\begin{equation}
C(\theta,d)=\bigl\lvert O_d^{n_{\text{max}},0}(\theta)\bigr\rvert=\bigl\lvert O_d^{0,m_{\text{max}}}(\theta)\bigr\rvert = \left\lvert\sqrt{\binom{d}{q_{\theta,d}}}(\cos\theta)^{d} (\tan\theta)^{q_{\theta,d}}\right\rvert,
\label{supeq:C}
\end{equation}
where $q_{\theta,d}=\lfloor d\sin^2\theta-\cos^2\theta\rfloor + 1$.

The second step is to find how fast $C(\theta,d)$ goes to zero. For this task, the floor function might be omitted, thus $q_{\theta,d}\approx d\sin^2\theta+1-\cos^2\theta=d\sin^2\theta+\sin^2\theta=(d+1)\sin^2\theta\approx d\sin^2\theta$ for large $d$.  The binomial coefficient can be approximated with Stirling's formula $n!\approx (n/e)^n\sqrt{2\pi n}$ in the following way
\begin{equation}
\begin{aligned}[t]
\binom{d}{d\sin^2\theta} ={}& \dfrac{d!}{(d\sin^2\theta)!\,(d-d\sin^2\theta)!} = \dfrac{d!}{(d\sin^2\theta)!\,(d\cos^2\theta)!}
\\
{}\approx{}& \dfrac{(d/e)^d\sqrt{2\pi d}}{(d/e\sin^2\theta)^{d\sin^2\theta}\sqrt{2\pi d\sin^2\theta}(d/e\cos^2\theta)^{d\cos^2\theta}\sqrt{2\pi d\cos^2\theta}}
\\
{}={}& \dfrac{1}{\sqrt{2\pi d}\Bigl[(\sin^2\theta)^{\sin^2\theta}(\cos^2\theta)^{\cos^2\theta}\Bigr]^d\sin\theta\cos\theta}.
\end{aligned}
\end{equation}
Therefore,
\begin{equation}
\begin{aligned}[t]
\sqrt{\binom{d}{q_{\theta,d}}} \approx {}& \dfrac{1}{\sqrt[4]{2\pi d}\Bigl[(\sin\theta)^{\sin^2\theta}(\cos\theta)^{\cos^2\theta}\Bigr]^{d}\sqrt{\sin\theta\cos\theta}}.
\end{aligned}
\end{equation}
Additionally,
\begin{equation}
(\cos\theta)^{d} (\tan\theta)^{q_{\theta,d}}\approx (\cos\theta)^{d} (\tan\theta)^{d\sin^2\theta} = \Bigl[(\sin\theta)^{\sin^2\theta}(\cos\theta)^{\cos^2\theta}\Bigr]^{d}.
\end{equation}
This leads to the following formula
\begin{equation}
C(\theta,d)\approx  \dfrac{\Bigl[(\sin\theta)^{\sin^2\theta}(\cos\theta)^{\cos^2\theta}\Bigr]^{d}}{\sqrt[4]{2\pi d}\Bigl[(\sin\theta)^{\sin^2\theta}(\cos\theta)^{\cos^2\theta}\Bigr]^{d}\sqrt{\sin\theta\cos\theta}} = \dfrac{1}{\sqrt[4]{2\pi d}\sqrt{\sin\theta\cos\theta}} = \dfrac{1}{\sqrt[4]{\tfrac{\pi}{2} d}\sqrt{\sin2\theta}} \sim \dfrac{1}{\sqrt[4]{d}}.
\end{equation}

\section*{Finding optimal experimental settings}

Here we present our numerical method applied to find the experimental settings allowing to observe violation of the quantum steering inequality.  The goal is to compute the optimal number of settings $N_{\text{opt}}$ and the set of angles $\theta_x$ ($x=1,\ldots,N_{\text{opt}}$) and $\theta_i$ which would maximize $V_Q$ given by Eq.~(\ref{supeq:V_Q}).  Maximal value of $V_Q$ ensures that $V_Q^\eta$ also reaches its maximum.

\begin{figure}\centering
\includegraphics[height=5cm]{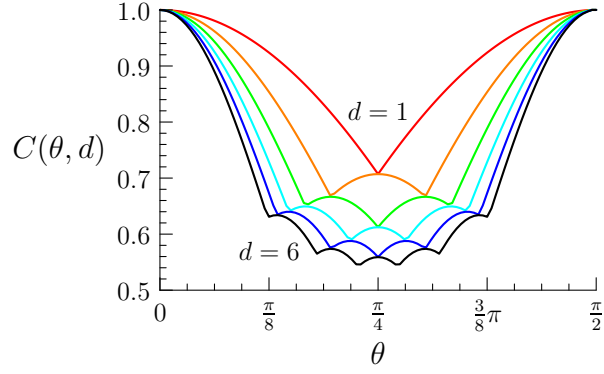}
\caption{$C(\theta,d)$ as a function of $\theta$ for $d=1,\ldots,6$.  Global minimum of this function is achieved for various angles.  The analysis is presented in the text.}
\label{supfig:C}
\end{figure}

The denominator of $V_Q$, $S_{\text{LHS}}$, depends on the sum of $C_i=C(\theta_i,d)$ which has to be minimized.  Therefore, our first step is based on analysis of $C(\theta,d)$ (Eq.~\ref{supeq:C}), depicted in Fig.~\ref{supfig:C} as a function of $\theta$ for $d=1,\ldots,6$.  This function is periodical with period $\tfrac{\pi}{2}$ and symmetrical with respect to $\theta=\tfrac{\pi}{4}$.  It always reaches $d$ minima, one or two of them being global.  In case of odd $d=1,3,5,\ldots$, the global minimum is achieved for $\theta=\tfrac{\pi}{4}$.  In contrast, even $d=2,4,6,\ldots$ regularly lead to two equal global minima, located symmetrically to the left and right of $\theta=\tfrac{\pi}{4}$.  Interestingly, all minima found for $d$ are always higher than minima for $d+1$ and lower than for $d-1$.  In case of odd $d$ we used $\theta_{x=1}<\theta_{x=2}<\ldots<\theta_{x=N-1}$ and $\theta_i=\theta{x=i+1}-\theta_{x=i}$.  We numerically optimized $\sum_{i=1}^{N-1}$ and found best $\theta_i=i\tfrac{\pi}{2N}$, thus $\theta=\min_{x,y}\lvert \theta_x-\theta_y\rvert=\tfrac{\pi}{2N}$.  This result was confirmed analytically for $d=1$ and $N=2$.  In case of even $d$ this set is not the best possible, since it may not include the global minima, but allows us to assure that $S_{\text{LHS}}(d+1)\leq S_{\text{LHS}}(d)$.  The maximal number of settings $N$ is limited by the number of settings of the stepper motor driving the polarizer, usually around 400 per full rotation.

In the second step our algorithm goes through all $N=1,2,\ldots$, computes $V_Q$ for given $d$ and finds the maximal value.  The result is a pair $\big(N_{\text{opt}}, V_Q\big)$.  The bounds for the lossy case are then computed as
\begin{equation}
  V_Q^{\eta} = \eta^{d}V_Q
\end{equation}
Figs.~4--5 in the main text display $V_Q$ and $V_Q^{\eta}$ computed with the above method.

It is worth noting that the erratic shape of $N_{\text{opt}}(d)$ is not accidental and results from Eq.~(\ref{supeq:C}).  Generally, the higher $d$, the more settings have to be used in order to maximize $V_Q$.  However, this relation is not monotonic and allows to find higher values of $d$ for which it is possible to use smaller $N_{\text{opt}}$.  This feature allows to better suit the limitations of the experimental setup.

\end{document}